\documentclass[reprint,aps, superscriptaddress,nofootinbib]{revtex4-1}

\usepackage{graphicx}
\usepackage{dcolumn}
\usepackage{bm}

\usepackage{amsmath}
\usepackage{amssymb}
\usepackage{amsfonts}
\usepackage{amsthm}
\usepackage{color}
\usepackage{MnSymbol}
\usepackage{caption}

\begin{document}

\title{Requirements for probing chiral Casimir-Polder forces in a molecular Talbot-Lau interferometer}

\date{\today}

\author{Fumika Suzuki}
\email{fsuzuki@lanl.gov}
 \affiliation{%
Theoretical Division, Los Alamos National Laboratory, Los Alamos, New Mexico 87545, USA
}
 \affiliation{%
Center for Nonlinear Studies, Los Alamos National Laboratory, Los Alamos, New Mexico 87545, USA
}

\author{S. A. Shah}

 \affiliation{%
Theoretical Division, Los Alamos National Laboratory, Los Alamos, New Mexico 87545, USA
}
 \affiliation{%
Center for Nonlinear Studies, Los Alamos National Laboratory, Los Alamos, New Mexico 87545, USA
}
\author{Diego A. R. Dalvit}
 \affiliation{%
Theoretical Division, Los Alamos National Laboratory, Los Alamos, New Mexico 87545, USA
}
\author{Markus Arndt}
\email{markus.arndt@univie.ac.at}
\affiliation{University of Vienna, Faculty of Physics\& VDSP\& VCQ, Boltzmanngasse5, A-1090 Vienna}

\begin{abstract}  
  We theoretically investigate the influence of chiral Casimir-Polder (CP) forces in Talbot-Lau interferometry, based on three nanomechanical gratings. We study scenarios where the second grating is either directly written into a chiral material or where the nanomask is coated with chiral substances. We show requirements for probing  enantiospecific effects in matter-wave interferometry in the transmission signal and the interference fringe visibility, which depend on the de Broglie wavelength  and the molecular chirality. The proposed setup is particularly sensitive to CP forces in the non-retarded regime where  chiral effects can be comparable in magnitude to their electric and magnetic counterparts.  While the first and third gratings do not change the phase of the matter wave, applying a coating of chiral substances to them  enhances the instrument's chiral selectivity.
\end{abstract}

\maketitle

\section{Introduction}
\captionsetup[figure]{justification=raggedright,singlelinecheck=false}
\captionsetup[table]{justification=raggedright,singlelinecheck=false}

Chirality-dependent dispersion forces, such as the van der Waals (vdW) forces, the Casimir-Polder (CP) forces and the Casimir effects have been theoretically predicted and analyzed in many prior studies, more recently also with a growing attention towards their experimental detection \cite{Casimir1948, chiral, chiral2, chiral3, chiral4, suzuki, chiral5}. 
A better understanding of these interactions promises deeper insight into quantum field theory \cite{chiral2}. Enantiospecific interactions are also important in the understanding of Hund's paradox \cite{hund,hund2,hund3} and of  homochirality, i.e., the question why many biomolecules in nature appear with only one chirality \cite{homo}. 

Recent studies of chirality span a diverse range of topics, including chirality-induced spin selectivity  \cite{CISS, CISS2}, proximity-induced chiral quantum light \cite{chirallight}, induced chirality in clusters \cite{ref3},  circular dichroism in  carbon nanotubes \cite{ref}, chiral AC Stark effects \cite{ac} and chiral resonance energy transfer \cite{energy}. Although enantiospecific \textit{optical }forces have  been observed experimentally \cite{optical,optical2,optical3,optical4,optical5},  the detection of chirality-dependent  \textit{dispersion} forces has remained a great challenge.

\begin{figure}
\includegraphics[clip,width=1\columnwidth]{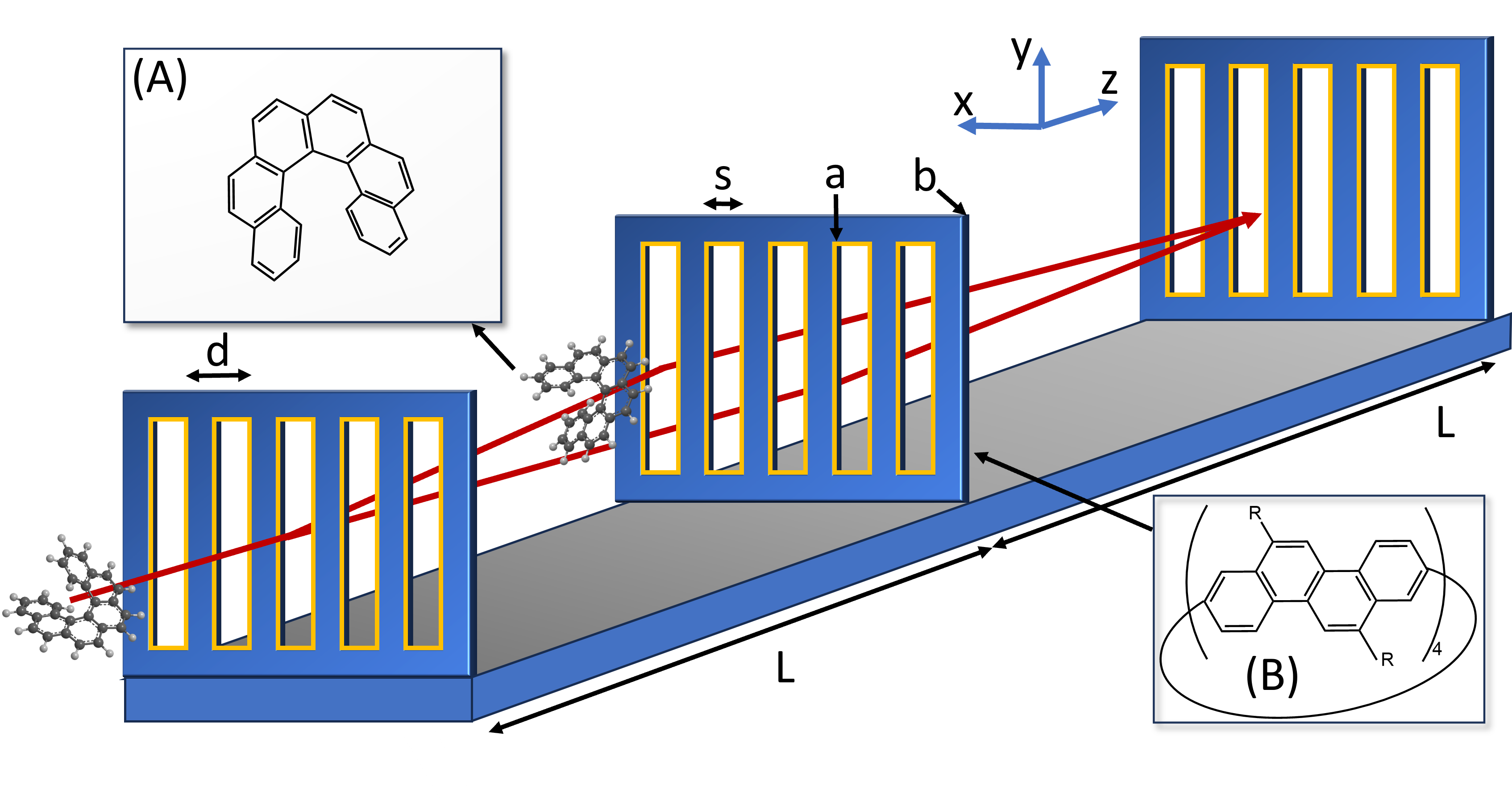} 
\caption{ A symmetric Talbot-Lau setup consists of three  gratings separated by equal distances $L$. In our present Gedanken-experiment, all nanomechanical gratings have a period  $d$, the thickness $b$ and an open fraction $f=s/d$, where $s$ is the geometrical slit width. The gratings can be made either from inherently chiral material or be written more conventionally into SiN$_\mathrm{x}$ and decorated by a layer of chiral substances  with thickness $a$. As an example, We show  hexahelicene  as the delocalized probe molecule (A) interacting with [4]cyclo-
2,8-chrysenylene (short name: [4]CC) as a chiral surface coating (B). Molecules with different properties are discussed in the text.}
\label{fig1}
\end{figure}

To enhance the  sensitivity to  {\it chiral} CP forces, \cite{suzuki} proposed to use  slow molecules in electronically excited states and separate  enantiomers in a planar chiral cavity. Excited states are appealing as they can exhibit strongly enhanced interaction strengths, but the short lifetime of dipole allowed transitions also limits the interaction time and the maximal momentum transfer and separation in molecular beam experiments.   
While the separation and analysis of enantiomers is commonplace in chemistry,  our following proposal relies on non-contact distant interactions in high vacuum, which are not commonly accessible.   
In our present work, we analyse in particular how the interaction of chiral molecules with {\it chiral} nanogratings  influences both the molecular transmission and the molecular fringe visibility in a near-field matter-wave interferometer (see Fig. 1).  

The CP potential can be understood as the vdW potential augmented by the influence of retardation \cite{Casimir1948}, and it has  already been discussed in the context of matter-wave diffraction of neutral atoms \cite{Grisenti1999, Bruehl2002,Lonij2010,Garcion2021,Bouton2023}, non-polar \cite{Arndt1999, Brezger2002, Brand2015} and polar molecules \cite{Knobloch2017}.  
The phase accumulated by a molecule in transit through a  grating depends on the molecular 
polarizability,  anisotropy, dipole moment, conformation,   the molecular 
velocity  as well as the grating slit width, thickness and shape. The dielectric function of the grating is relevant across all frequencies and one may also need to account for charge impurities in the grating~\cite{Knobloch2017}. 
The interaction with nanostructures is favorable as  {\em chiral} CP forces  decay faster with distance  than their electric and magnetic counterparts in the retarded regime.   We  propose specifically Talbot-Lau interferometry (TLI), as shown in Fig.~\ref{fig1}, 
as a suitable tool to explore proximity effects since TLI can create coherent self-images of several thousand molecular beams in parallel, all with a low velocity normal to the grating wall. 

In the following, we study the conditions under which chiral molecules of sufficiently high rotatory strength may be discriminated by the interferometer transmission and fringe visibility.   The paper is organized as follows: In Sec. II, we review the theoretical framework to study the influence of CP forces in Talbot-Lau interferometry with three gratings. In Sec. III, we investigate the impact of chiral CP forces when the second grating $G_2$ is perfectly chiral and show requirements for observing differences in interferometer transmission and fringe visibility between the enantiomers. In Sec. IV, we study a more realistic scenario where the second grating is inherently non-chiral but coated with chiral substances of high rotatory strength. In Sec. V, we then assume all gratings to be equal and find that chiral coatings of gratings $G_1$ and $G_3$ have a  strong influence on the enantiomer discriminating power of the instrument, as the chiral CP forces modify the grating transmission function substantially. This can result in observable differences between enantiomers in  the total signal strength. Finally, we discuss experimental considerations in Sec. VI and conclude in Sec. VII.

\section{CP forces in Talbot-Lau interferometer}
We begin by presenting an overview of the framework employed for studying the effects of CP forces on chiral molecules  in a Talbot-Lau interferometer (Fig.~\ref{fig1}) composed of three nanogratings, G1-G3, each with a  thickness of  $b$ and  a periodicity of $d$. We choose a grating distance of $L$. The grating slit width $s$  may be augmented by a chiral  coating of thickness $a$. It can be also reduced by the forces to a cutoff $x_c$,  discussed below. Note that such an interferometer can also be operated with only two gratings, if G3 is replaced by a molecular binding surface, which can be imaged with 10\,nm resolution and high efficiency~\cite{Juffmann2009}. Even G1 could be replaced by a point source.  

We start by summarizing the  expressions for the expected interferometer signal and fringe visibility and will then add the effect of chiral potentials to the result. This section  uses Wigner functions in phase space as described in~\cite{klaus}:
\begin{eqnarray}
w (\mathbf{r},\mathbf{p})=\frac{1}{(2\pi \hbar)^2}\int d\mathbf{\Delta}e^{i\mathbf{p}\mathbf{\Delta}/\hbar}\rho\left(\mathbf{r}-\frac{\mathbf{\Delta}}{2},\mathbf{r}+\frac{\mathbf{\Delta}}{2}\right)
\end{eqnarray}
where $\rho(\mathbf{r},\mathbf{r'})$ is the position density matrix and $\mathbf{\Delta}=\mathbf{r}-\mathbf{r}'$.

In a Talbot-Lau setup, the molecular beam traverses three gratings $G_1-G_3$, spaced at equal distances $L$. The Wigner function after the first grating can be expressed as
\begin{eqnarray}
w_1(\mathbf{r},\mathbf{p})=|t_1(\mathbf{r})|^2,
\end{eqnarray}
assuming that the initial Wigner function is normalized and homogeneous, $w_0(\mathbf{r},\mathbf{p})=1$, i.e., the molecules enter the grating without any prior spatial coherence and evenly from all directions. We further assume a symmetric interferometer where all gratings have the same period $d$ and distance $L$.  
The geometrical transmission function can be written as the Fourier series $t_1(x)=\sum_l a_{l} \exp(2\pi i l x/d)$, where the coefficients $a_l$ depend on the slit width, slit thickness, coating and also on the interaction between the molecules and the grating wall. We assume a monochromatic molecular beam and  will analyze the dependence of the interference fringes on the velocity, i.e., on the de Broglie wavelength.

A free evolution of the molecule over a distance $L$ between the gratings changes the Wigner function
\begin{eqnarray}
w_1(\mathbf{r},\mathbf{p}) \rightarrow w_1(\mathbf{r}-L\mathbf{p}/p_z,\mathbf{p}),
\end{eqnarray}
where $p_z$ is the momentum of the molecule in $z$-direction.  We neglect all external accelerations due to gravity, assuming that all gratings are aligned to better than 100\,$\mu$rad with the line of gravity. We also neglect the Coriolis acceleration $a_c=2v \times \Omega_E$ which is about thousand times smaller than the gravitational acceleration for molecules travelling at about 200\,m/s, when $\Omega_E=72$\,µrad/s is the frequency of the rotation of the Earth and can be compensated by gravity under suitable interferometer alignment \cite{Fein2020}.  

The passage through each second grating modulates the wavefunction according to $\psi_1 (\mathbf{r})\rightarrow t_2(\mathbf{r})\psi_1(\mathbf{r})$ where $t_2(\mathbf{r})$ encodes the complex transmission function. 
The Wigner function after passage through a second grating is thus 
\begin{eqnarray}
w_1 (\mathbf{r},\mathbf{p}) \rightarrow \int d\mathbf{q}\,T (\mathbf{r},\mathbf{q}) w_1(\mathbf{r},\mathbf{p}-\mathbf{q})
\end{eqnarray}
where
\begin{eqnarray}
T(\mathbf{r},\mathbf{p})=\frac{1}{(2\pi\hbar)^2}\int d\mathbf{\Delta} e^{i\mathbf{p}\mathbf{\Delta}/\hbar}t^{*}_2\left(\mathbf{r}+\frac{\mathbf{\Delta}}{2}\right)t_2\left(\mathbf{r}-\frac{\mathbf{\Delta}}{2}\right).\end{eqnarray}

Talbot-Lau interferometers can generate spatial coherence even from an initially incoherent molecular beam by diffracting the molecules at the slits of the first grating, if the first grating is  an absorptive, spatially selective mask -- such as a nanomechanical structure or a photo-depletion grating \cite{Nimmrichter2011}. Diffraction within each slit of grating $G_1$ then prepares the transverse coherence required to achieve spatial superposition and diffraction at $G_2$. While we can treat $G_1$ as a pure transmission filter, its CP interaction with the molecule must still be treated carefully. Molecular trajectories are deflected to undetected regions if they pass very close by the grating walls. The effective slit width is thus narrowed by a cut-off distance $2x_c^{(1)}$ which influences both the interference contrast and the transmitted signal. Furthermore, this effect is dispersive: slower molecules are even further deflected and do no longer contribute to the signal.    

The second grating $G_2$ then modifies both the amplitude and the phase of the already delocalized matter-wave. In eikonal approximation the transmission function of the second grating reads:
\begin{eqnarray}
t_2 (x)\rightarrow t_2 (x)\exp \left(-i\frac{m b}{p_z}\frac{V(x)}{\hbar}\right)
\end{eqnarray}
where $m$ is the mass of the molecule. Note that this additional phase depending on $V(x)$ is canceled out for G1 and G3, leaving only the influence of a cutoff induced by the CP force, as discussed below.  The quantum coherence term only exists for waves traversing through G2. Nonetheless, the contribution of  G1 and G3 to the final signal and visibility is relevant as they modulate the effective transmission. 

Assuming that the second grating has the Fourier coefficients $b'_m$, we find
\begin{eqnarray}
T(x,p)=\int dq \,T_0 (x,p-q) T_V (x,q)
\end{eqnarray}
where
\begin{eqnarray}
T_0(x,p)&=&\displaystyle\sum_{l,j\in\mathbb{Z}}b'_j b'^{*}_{j-l}\exp \left(2\pi i l \frac{x}{d}\right)\delta \left(p-\hbar\pi\frac{2j-l}{d}\right),\nonumber\\
T_V (x,q)&=&\frac{1}{2\pi \hbar}\int d\Delta e^{iq\Delta/\hbar}\nonumber\\
&\times& \exp \left(-i\frac{mb}{p_z \hbar}\left[V\left(x-\frac{\Delta}{2}\right)-V\left(x+\frac{\Delta}{2}\right)\right]\right).
\end{eqnarray}

After free evolution over the distance $L$, passage through grating $G_2$ and another free evolution over the distance $L$, the Wigner function reads
\begin{eqnarray}
w (\mathbf{r},\mathbf{p})=\int d\mathbf{q}\left| t_1 \left(\mathbf{r}-2L\frac{\mathbf{p}}{p_z}+\frac{\mathbf{q}}{p_z}L\right)\right|^2 T\left(\mathbf{r}-\frac{\mathbf{p}}{p_z}L,\mathbf{q}\right).\nonumber\\
\end{eqnarray}

We obtain the molecule density by integrating the Wigner function over all momenta and obtain
\begin{eqnarray}
w(x) \propto \displaystyle\sum_{l\in \mathbb{Z}} A_l^{*} B_{2l}^{(\lambda)}\exp \left(2\pi i l\frac{x}{d}\right).
\end{eqnarray}
Here, the Talbot coefficients of the first   grating are 
\begin{equation}
    A_l =\displaystyle\sum_{j\in \mathbb{Z}}a_j a_{j-l}^{*}
\end{equation}
while the coefficients of the second grating are  
\begin{equation}
B_l^{(\lambda)}=\displaystyle\sum_{j\in\mathbb{Z}}b_j b^{*}_{j-l}\exp \left(i\pi \frac{l^2-2jl}{2}\frac{L}{L_{\lambda}}\right).
\end{equation}
Here we have introduced the Talbot length as $L_{\lambda}=d^2/\lambda$ and the de Broglie wavelength as $\lambda=h/p_z$. 

The Talbot coefficients above are modified by the CP potential, 
\begin{equation}
b_l=\sum_j b'_j c_{l-j}
\end{equation}
where
\begin{equation}    
\label{ccoeff}
c_l =\frac{1}{d}\int_{-d/2}^{d/2}e^{-2\pi i lx/d}\exp \left(-i\frac{mb}{p_z}\frac{V(x)}{\hbar}\right) dx.
\end{equation}
This includes the chiral CP potential $V (x)$, which will be discussed below.

While the molecular interference pattern can be captured and imaged on a surface~\cite{Juffmann2009,Juffmann2012}, for non-fluorescent molecules it has become common to use the spatial resolution of a third transmission mask $G_3$. The transmitted molecular flux behind $G_3$  is then a convolution of the density $w(x)$ before the grating and the grating intensity transmission function $|t_3(x)|$:
\begin{equation}
    S(x_3)=\int w (x) |t_3 (x-x_3)|^2 dx.
\end{equation}

It then suffices to add a mass-selective large molecule counter~\cite{Hackermuller2003, Strauss2023}. The expected signal is then \cite{klaus}
\begin{equation}\label{signaleq}
    S(x_3) \propto \displaystyle\sum_{l\in\mathbb{Z}}A_l^{*}A_l'^{*}B_{2l}^{(\lambda)}\exp\left(2\pi i l\frac{x_3}{d}\right)
\end{equation}\label{signal}where $A_l'$ are the Talbot coefficients of the third grating and we extract the sinusoidal fringe visibility
\begin{equation}\label{vis}
\mathcal{V}=\frac{S_{\rm max}-S_{\rm min}}{S_{\rm max}+S_{\rm min}}=\frac{\left|\displaystyle\sum_{n=1}^{\infty}A_{2n-1}A'_{2n-1}B_{4n-2}^{(\lambda)}\right|}{\frac12 A_0A_0' B_0^{(\lambda)}+\displaystyle\sum_{n=1}^{\infty}A_{2n}A'_{2n}B_{4n}^{(\lambda)}}.
\end{equation}

In the following, we employ these equations to explore the differences in signal and fringe visibility between molecular enantiomers in the presence of chiral CP forces.

Note that  molecular orientation is generally relevant in diffraction at nanomechanical masks, since it exposes different components of the polarizability tensor to the grating surface. We  assume the chirality to be at a fixed temperature and we average over the molecular orientations.   

\section{Second grating written into a chiral material}
We start by investigating a scenario in which the second grating is made from a chiral material, characterized by the reflection matrix:
\begin{eqnarray}
\mathbf{R}=\begin{pmatrix}r_{ss}&&r_{sp}\\r_{ps}&&r_{pp}\end{pmatrix}=\begin{pmatrix}-r&&r_c\\r_c&&r\end{pmatrix}.
\end{eqnarray}
It describes how  the $p$- and $s$-polarized components of linearly polarized light, $\mathbf{E}_s$, $\mathbf{E}_p$ are reflected by the chiral material: $\mathbf{E}_s \rightarrow r_{sp}\mathbf{E}_p +r_{ss}\mathbf{E}_s$  and $\mathbf{E}_p\rightarrow r_{pp} \mathbf{E}_{p}+r_{ps} \mathbf{E}_{s}$. For a non-chiral surface, $r_{sp}=r_{ps}=0$ \cite{isotropic}.

The CP potential between the grating and a chiral molecule in the ground state can be expressed as the sum of its electric, magnetic, and chiral components $V=V_{e}+V_{m}+V_{c}$, where \cite{chiral3, scheel, stefan}
\begin{eqnarray}\label{potentials}
V_{e} & = & \frac{\hbar \mu_0}{2\pi}\int_0^{\infty} d\xi \, \alpha (i\xi) \xi^2  \mbox{tr}\mathbf{G} (\mathbf{r},\mathbf{r},i\xi), \nonumber \\
V_{m} & = &\frac{\hbar\mu_0}{2\pi}\int_0^{\infty}d\xi \, \beta (i\xi)\mbox{tr}[\nabla \times \mathbf{G}(\mathbf{r},\mathbf{r},i\xi)\times \stackrel{\leftarrow\mbox{ }}{\nabla'}], \nonumber \\
V_{c}&=&-\frac{\hbar\mu_0}{\pi}\int_0^{\infty}d\xi \, \Gamma (i\xi) \xi  \mbox{tr}[\nabla \times \mathbf{G}(\mathbf{r},\mathbf{r},i\xi)]
\end{eqnarray}
and
\begin{eqnarray}\label{alpha}
\alpha (i\xi) &=&\frac{2}{3\hbar}\sum_k \frac{\omega_{k}|\mathbf{d}_{0k}|^2}{\omega_{k}^2+\xi^2},\quad \beta (i\xi) =\frac{2}{3\hbar}\sum_k \frac{\omega_{k}|\mathbf{m}_{0k}|^2}{\omega_{k}^2+\xi^2},\nonumber\\
\Gamma (i\xi)&=&-\frac{2}{3\hbar}\sum_k \frac{\xi R_{0k}}{\omega^2_{k}+\xi^2}.
\end{eqnarray}
Here $\nabla$ and $\stackrel{\leftarrow\mbox{ }}{\nabla'}$ act on the first and second position arguments of the Green’s tensor only, $\mu_0$ is the vacuum permeability, $\omega_{k}$ the transition frequency, $\mathbf{d}_{0k}$ is the electric dipole transition matrix element, $\mathbf{m}_{k0}$ is the  magnetic dipole matrix element, and $R_{0k}=\mbox{Im}(\mathbf{d}_{0k}\cdot \mathbf{m}_{k0})$  is the optical rotatory strength. Formally, the sum is over all optical transitions (see Eq. \ref{alpha}) but we can limit the computations to the dominant  transition (i.e., $0\rightarrow k=1$). Both enantiomers have approximately the same modulus but opposite signs of the rotatory strength $R_{0k}$ and the distinct interference patterns observed between the two enantiomers can be attributed to the different signs of their chiral potential $V_c$. 

The Green's tensor $\mathbf{G}(\mathbf{r},\mathbf{r}, \omega)$ includes the relevant information about the scattering of photons at the  grating  walls and it is written as \cite{green}
\begin{eqnarray}\label{greens}
&&\mathbf{G}(\mathbf{r},\mathbf{r},\omega)\nonumber\\
&&=\frac{i}{8\pi^2}\int\frac{d^2k^{\parallel}}{k^{\perp}}e^{2ik^{\perp}x}\sum_{\sigma_1\sigma_2 =s,p}r_{\sigma_1\sigma_2}\mathbf{e}_{\sigma_1}(k^{\perp})\mathbf{e}_{\sigma_2}(-k^{\perp})\nonumber\\
\end{eqnarray}
where $k^{\parallel}$ and $k^{\perp}$ are the components of the wave vector parallel and perpendicular to the grating surface and $k^{\perp2}=\omega^2/c^2-k^{\parallel 2}$. Substituting (\ref{alpha}) and (\ref{greens}) into (\ref{potentials}), we obtain \cite{chiral3, scheel, stefan}
\begin{eqnarray}\label{pot}
V_e (x) &=&-\frac{r |\mathbf{d}_{01}|^2}{48\pi \epsilon_0 x^3}, \nonumber\\ V_{m}(x) &=&\frac{r|\mathbf{m}_{01}|^2}{48\pi \epsilon_0 c^2 x^3},\nonumber\\
V_c (x) &=&\frac{ r_c \mu_0 c }{12\pi^2 x^3} R_{01}\log (\omega_{1} x/c),
\end{eqnarray}
where $\epsilon_0$ is the vacuum permittivity and $x$ is the  distance between the molecule and the grating wall.  Since we are considering the opening of the grating slit to be around 100~nm in size, we employed the non-retarded limit approximation $x\omega_1 /c \ll 1$, which is favorable for the chiral CP potential. In the retarded regime ($x\omega_{1} /c \gg 1$) the chiral potential ($V_c\propto 1/x^5$)  decays faster with distance than its electric and magnetic counterparts ($V_{e,m}\propto 1/x^4$) \cite{chiral3}. Since the molecule is significantly smaller than the grating thickness, we also use the proximity force approximation, wherein a molecule is assumed to interact with the grating like with an infinitely extended surface. 
The approach and departure of the molecule, i.e., forces acting before and after grating can modify the interference fringes. However, even after including such effects in a careful theory using a Green's function approach~\cite{nano}, every grating model had to be corrected by a fudge factor in the Casimir force. This correction factor shrunk, however, from $\eta\simeq 8/5/2$  for  gratings with a period of d=100\,nm  and a thickness of $t=12/45/87$\,nm. It should be negligible in our case with  $t=160$\,nm.  
Moreover, all fringe effects and effects of defects in the grating are stationary and even in their presence one can see  differences of interactions with molecules of different properties. This has also recently been exploited in nano-grating interference of molecules with and without electric dipole moments~\cite{Simonovic2024}. In our case, the influence of chirality can then be probed differentially by swapping between the enantiomers.

Molecules close to the grating wall are  attracted by the dispersive force. If they come within a distance shorter than a cut-off distance $x_c$, which depends on the materials and their chirality, they may be deflected to beyond the acceptance angle of the detector or even  collide with the  grating wall. In a typical Talbot Lau setup, the cut-off distance for the first and second grating $x_c^{(1)}$ and $x_c^{(2)}$ can be estimated to be
\begin{eqnarray}\label{cutoff0}
\frac{\Delta p_x}{p_z}=\frac{F (x_c^{(1),(2)})\Delta t}{p_z}=\frac{F (x_c^{(1),(2)})m b}{p_z^2}=\theta
\end{eqnarray}
where $\theta=1$\,mrad and 2\,mrad are common experimental values at the first and second grating, respectively~\cite{Brezger2002}. Here, the force $F(x)=-V'(x)$ is the gradient of the CP potential.
For the third grating, $x_c^{(3)}$ is given by \cite{cutoff,cutoff2}
\begin{eqnarray}\label{cutoff}
\frac{b}{v_z}=\sqrt{\frac{m}{2}}\int_0^{x_c^{(3)}}dx\frac{1}{\sqrt{-V(x)}}
\end{eqnarray}
where   $v_z$ is the longitudinal velocity of the molecule. The chirality dependence of the cut-off distance $x_c$ results in  different effective slit openings for different enantiomers. This influences both the interferometer fringe visibility and even more its transmission amplitude.

Since a large optical rotatory strength $R_{01}$ and a large value of $r_c$  result in a strong chiral CP force (\ref{pot}), we consider a beam of molecules whose optical rotatory strength is large compared to their electric and magnetic dipole moment. For the sake of the argument we  start by assuming $G2$ to be made from a perfectly chiral material. In theory such a structure is conceivable  even though it is hard to realize in the lab. In that case,  $r_c \rightarrow 1$ and $r \rightarrow 0$, which implies that $V_e$ and $V_m$  vanish and that only the chirality-dependent $V_c$  contributes. A  molecule with a positive value of $R_{01}$ is referred to as right-handed and it is attracted to the grating. Its left-handed enantiomer with $R_{01}<0$ is repelled and  only  molecules with  $R_{01}>0$  experience a chirality dependent cut-off  $x_c$ at a grating of positive chirality. 

We therefore expect a  difference between two enantiomers in their interferometer transmission and fringe contrast when we include the effect of the chiral CP potential,
\begin{eqnarray}    
    V (x)&=&     V_c (x+x_o)+V_c (x_o-x),
\end{eqnarray}
where we  let the grating walls be located at $\pm x_o$ with $x_o=fd/2$, and the molecules fly through them, i.e., $x\in (-x_o,x_o)$

 Substituting $V(x)$ into (\ref{ccoeff}) and performing the integration including the cut-off $x_c$,
 we obtain  Fig.\,2.  Here we let the grating thickness be $b=160\,$nm, the periodicity $d=257\,$nm, the grating distance $L=50\,$mm, and the open fraction $f=0.45$. These parameters are chosen close to non-chiral nanostructures used in earlier experiments \cite{Juffmann2009}.
 The top panel shows one period of the matter-wave fringe, i.e., the transmission signal behind $G3$ as a function of the position of the third grating. This is done both for right-handed (solid red line) and left-handed (dashed blue line) molecules. In the left panel (i), the simulation is performed for hexahelicene with a monochromatic velocity of $v_z = 180$\,m/s.  Their mass is  $m=328$\,Da,  and a rotatory strength of $|R_{01}|=700 \times 10^{-40}$\,cgs has been reported with a   transition frequency of $\omega_{1}=2\pi\times 10^{15}\,$s$^{-1}$ \cite{hexadata}. From such curves we can extract the interference fringe visibility which we display in the  bottom panel as a function of  the  molecular velocity (bottom scale) and de Broglie wavelength (top scale). The plots show that chirality of hexahelicene can be distinguished in the transmission signal. The panels of column (ii) depict the simulation for a hypothetical molecule with a mass $m=1000$\,Da and a rotatory strength 10 times greater than that of hexahelicene with same transition frequency. For the evaluation of the transmission signal, we set $v_z=140$\,m/s. This simulation demonstrates that, with such molecules, both the transmission signal and visibility as functions of velocity, could be differentiated for enantiomers due to the chiral CP forces, even though in a small velocity band. 
 
 The figure is the result of numerical computations based on Eq. (11) to Eq. (17), which is in agreement with the expectation that the final signal and fringe visibility must be periodic in the Talbot Length and thus the de Broglie wavelength. Electromagnetic properties of the molecules determine an effective open slit width of the grating, which enter the above equations. Due to chiral CP forces, enantiomers perceive different effective forces with the gratings, resulting in slight disparities in signal and visibility even for identical de Broglie wavelengths. This serves as the primary cause of the observed differences between enantiomers, as depicted in the figure.
 
Note that the assumption so far is strongly idealized: real-world materials generally have $r\neq 0$, and we require that the 
electric and magnetic anisotropy factor \cite{ani} $g_d:=|R_{01}/c|\,/\,|\mathbf{d}_{01}|^2$ and  $g_m:=|R_{01}\cdot c|\,/\,|\mathbf{m}_{01}|^2$ are sufficiently large to ensure  the chiral CP potential to be significant in comparison to its electric and magnetic counterparts (\ref{pot}):
\begin{equation}
r/r_c \leq g_e, g_m.    
\label{condition}
\end{equation}
A possible solution is  to use CP forces with opposite signs for the electric and magnetic components  as observed over a perfectly conducting plate \cite{scheel}. When the magnitudes of $\mathbf{d}_{01}$ and $\mathbf{m}_{01}/c$ are comparable, these two components may cancel, leaving only the chiral component. 

Our calculations and similar ones in the literature \cite{chiral,chiral3,suzuki}  rely on an effective-medium approximation, where the spatial variation of permittivity and permeability are neglected. This is only justified when the distance of the molecule  to the chiral material exceeds the size of the chiral structure of the metamaterial  \cite{diego}. This requires nanoscale artefacts inside the grating bars, which is technologically very demanding. To overcome this challenge,  we investigate  a scenario where the second grating is coated with chiral molecules in the following section.

\begin{figure}
\includegraphics[clip,width=1\columnwidth]{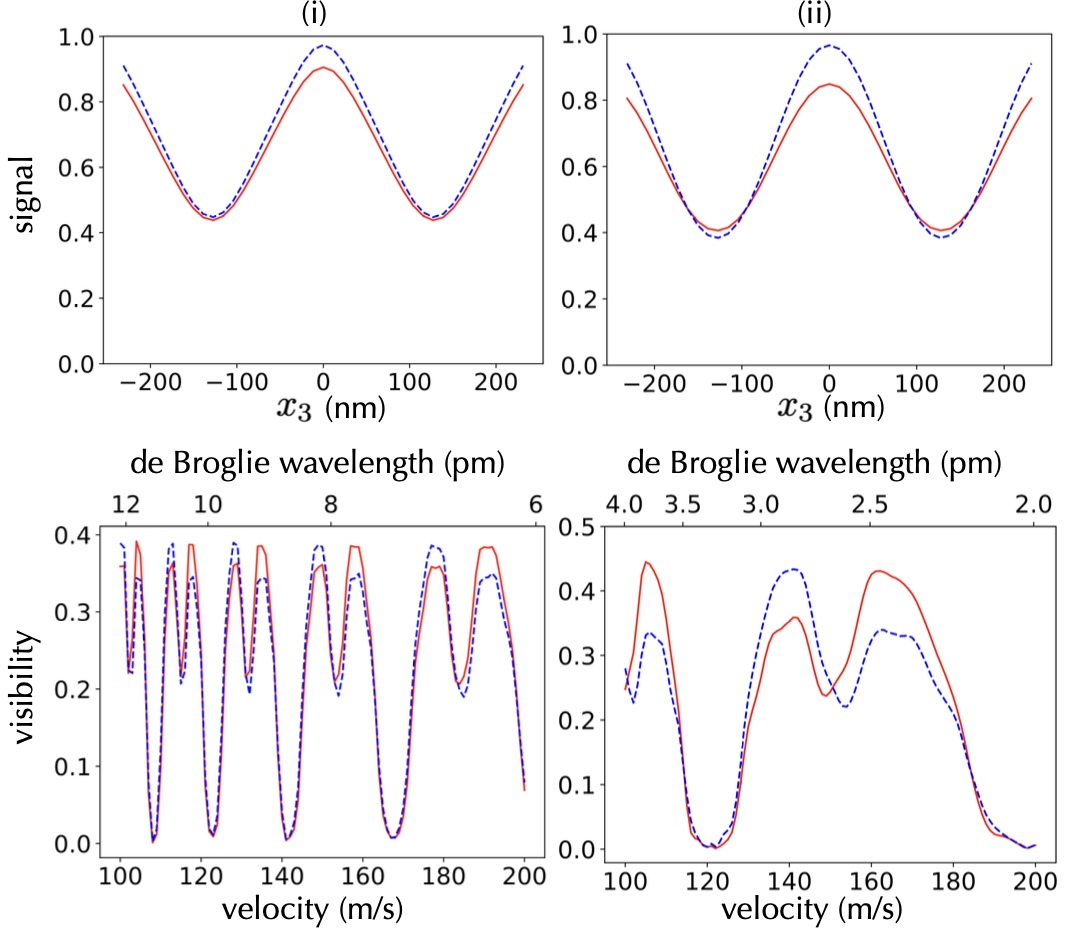}
\caption{ Predicted interferometer transmission as a function of the position of G3 (top) and fringe visibility as a function of molecular velocity (bottom). We compare  right-handed (solid red line) and left-handed (dashed blue line) molecules.  The simulation is done for (i) hexahelicene  and  (ii)  for a hypothetical molecule with $m=1000$\,Da and a rotatory strength ten times greater than that of hexahelicene. Here, the gratings $G_1$ and $G_3$ are assumed to be non-chiral while $G_2$ is assumed to be perfectly right-handed chiral.}  
\label{fig2}
\end{figure}

\section{Second grating coated with chiral molecules}
The electric, magnetic, and chiral components of the CP potential between two molecules can be written as \cite{stefan, andrews, salam}
\begin{eqnarray}
V_{e}(x)&=&-\frac{1}{24\pi^2 \epsilon_0^2x^6}\displaystyle\sum_{k,k'}\frac{|\mathbf{d}_{0k}^A|^2 |\mathbf{d}_{0k'}^B|^2}{E_{k}^A+E_{k'}^B},\nonumber\\
V_{m}(x)&=&-\frac{1}{24\pi^2 \epsilon_0^2 c^2x^6}\displaystyle\sum_{k,k'}\frac{|\mathbf{m}_{0k}^A|^2 |\mathbf{m}_{0k'}^B|^2}{E_{k}^A+E_{k'}^B},\nonumber\\
V_{c}(x)&=&-\frac{1}{12\pi^2 \epsilon_0^2 c^2x^6}\displaystyle\sum_{k,k'}\frac{R_{0k}^AR_{0k'}^B}{E_{k}^A+E_{k'}^B},
\end{eqnarray}
where  the electric dipole transition matrix elements, the magnetic dipole moment matrix elements, the optical rotatory strengths, and the energy differences between the $|0\rangle$,  $|k\rangle$ or $|k'\rangle$-state for molecule $A$ and molecule $B$ are denoted by $\mathbf{d}_{0k}^A$,  $\mathbf{d}_{0k'}^B$, $\mathbf{m}_{0k}^A$, $\mathbf{m}_{0k'}^B$, $R_{0k}^A$, $R_{0k'}^B$,  $E_{k}^A$, $E_{k'}^B$, respectively. Here, $A$ refers to the molecules propagating as  matter waves, while $B$ represents the molecules coating the second grating G2. 

By integrating over all $B$-molecules in a layer of thickness $a$ and number density $n_{B}$, we obtain the potential exerted by the coating:
\begin{eqnarray}\label{coated}
V_{\rm coat}(x)&=&-\frac{n_B}{144\pi\epsilon_0^2}\left(\frac{1}{(x-a)^3}-\frac{1}{x^3}\right)\displaystyle\sum_{m,n}\frac{|\mathbf{d}_{0m}^A|^2 |\mathbf{d}_{0n}^B|^2}{E_{m0}^A+E_{n0}^B}\nonumber\\
&&-\frac{n_B}{144\pi \epsilon_0^2 c^2}\left(\frac{1}{(x-a)^3}-\frac{1}{x^3}\right)\displaystyle\sum_{m,n}\frac{|\mathbf{m}_{0m}^A|^2 |\mathbf{m}_{0n}^B|^2}{E_{m0}^A+E_{n0}^B}\nonumber\\
&&- \frac{n_B}{72\pi \epsilon_0^2 c^2}\left(\frac{1}{(x-a)^3}-\frac{1}{x^3}\right)\displaystyle\sum_{m,n}\frac{R_{0m}^AR_{0n}^B}{E_{m0}^A+E_{n0}^B}
\end{eqnarray}
where $x$ is the distance between the molecule and the surface of the grating.

\begin{figure}
\includegraphics[clip,width=1\columnwidth]{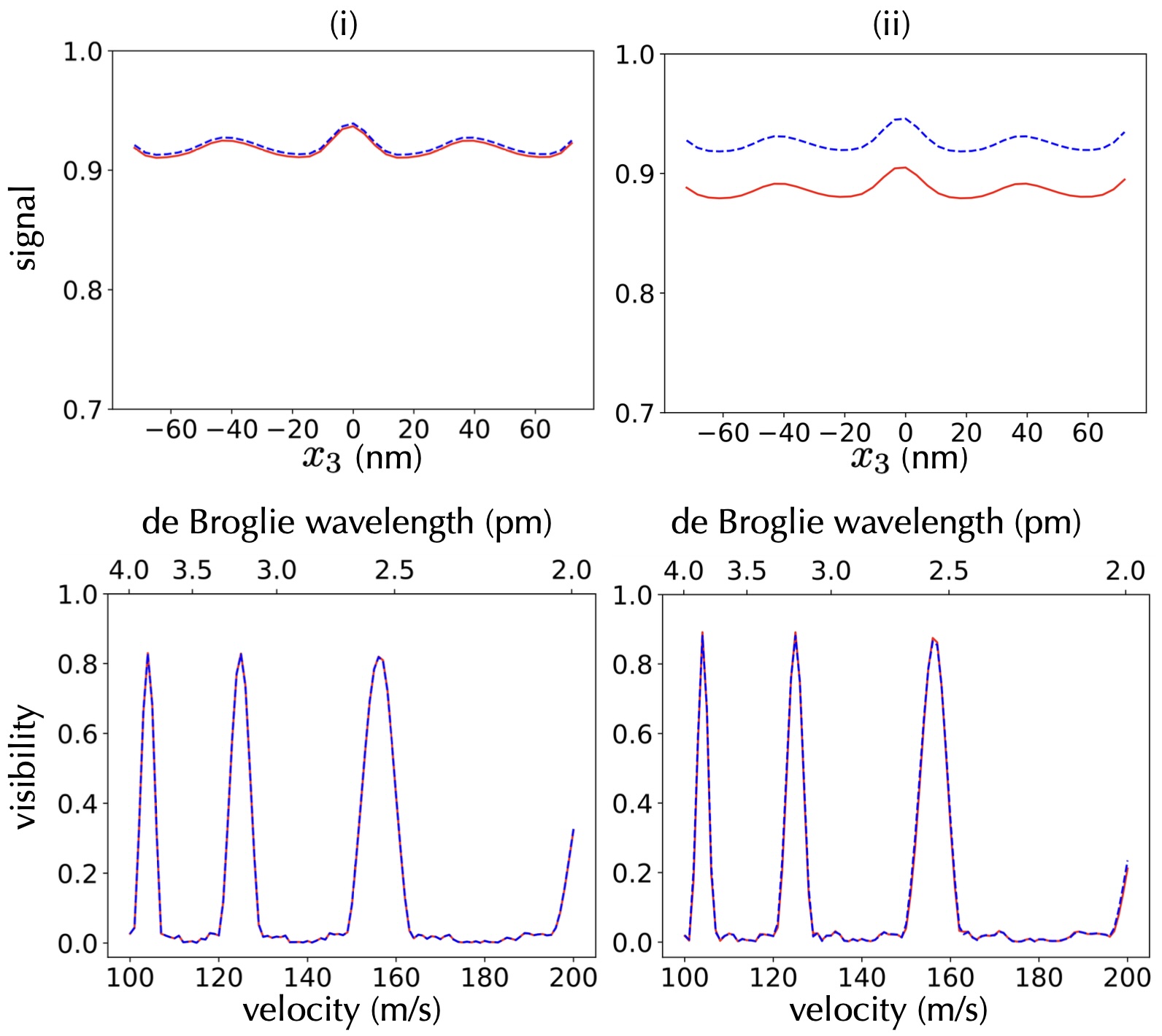} 
\caption{Interferometer transmission (top) and fringe visibility (bottom) compared for right-handed (solid red line) and left-handed (dashed blue line). The simulation is done for  (i) molecules with $|R_{01}|=1000\times 10^{-40}\,$cgs, $g_e=0.2$, $g_m=5$  and  (ii) molecules  with $|R_{01}|=5000\times 10^{-40}\,$cgs, $g_e=0.3$, $g_m=3.3$. In both cases, $m=1000\,$Da and $\omega_1=2\pi \times 10^{15}\,$s$^{-1}$. In this more realistic scenario, the grating substrates are all inherently non-chiral, but $G_2$ is  assumed to be coated by  chiral molecules with a layer thickness $a=10$\,nm. The velocity $v_z=140$\,m/s in the top panels.}
\label{fig3}
\end{figure}

In addition, we include the CP potential arising from the bare grating itself:
\begin{equation}\label{pot2}
V_{\rm grating}(x)=-\frac{\hbar}{16\pi^2 \epsilon_0 x^3}\int_0^{\infty}d\xi \alpha^A (i\xi) \frac{\epsilon (i\xi)-1}{\epsilon (i\xi)+1}
\end{equation}
where $ \alpha^A (i\xi)$ is given by (\ref{alpha}) and
\begin{equation}
\epsilon(\omega) =\frac{\Omega_L^2-\omega^2-i\omega\gamma_L}{\Omega_T^2-\omega^2-i\omega\gamma_T}.
\end{equation}
We consider a grating made of silicon nitride, where the resonance frequency of longitudinal mode $\Omega_L=2.69\times10^{16}$\,rad/s, the resonance frequency of the transverse mode $\Omega_T=1.33\times 10^{16}$\,rad/s, the longitudinal mode damping $\gamma_L=3.05\times10^{16}$\,rad/s, and the transverse mode damping $\gamma_T=6.40\times 10^{15}$\,rad/s \cite{SiN}.

To obtain the effect of the CP potentials on transmission signal and fringe visibility, we substitute 
\begin{eqnarray}
    V(x)&=&V_{\rm grating} (x+x_o)+V_{\rm grating} (x_o-x) \nonumber\\ 
    &+&V_{\rm coat}(x+x_o)+V_{\rm coat}(x_o-x), 
\end{eqnarray}
into (\ref{ccoeff}), and compute (\ref{signaleq}) and (\ref{vis}). The cut-off distance $x_c$ can again be evaluated using (\ref{cutoff0}) and (\ref{cutoff}).

Due to the limited availability of data regarding rotatory strength, electric, and magnetic anisotropy factors of molecules in chemistry, we base our  following simulations on a reasonable estimate of molecular properties to see the trends. Since the CP force from the coating is generally weak compared to that from the grating itself, we proceed under the assumption that the grating is characterized by a thickness of grating $b = 160\,$nm, a reduced periodicity $d = 80\,$nm, a shorter grating distance $L = 10\,$mm, an open fraction $f = 0.45$ and coating thickness of  $a=10\,$nm.  

Fig.\,\ref{fig3} (i) depicts the simulation where   $|R_{01}|=1000\times 10^{-40}\,$cgs, $g_e=0.2$ and $g_m=5$ for both matter-wave molecules and molecules coating the second grating G2. For (ii), it is assumed that $|R_{01}|=5000\times 10^{-40}\,$cgs, $g_e=0.3$ and $g_m=3.3$ for both  molecules. We set the transition frequency $\omega_1=2\pi \times 10^{15}\,$s$^{-1}$, the number density of the coating molecules $n_B=5\times10^{28}$/m$^3$ and $m=1000\,$Da for the molecules travelling as matter-waves.  The second grating is always coated with right-handed molecules. As depicted in the figure, we only observe a small difference between both enantiomers in the transmission signal. This is  hard to measure in an experiment and  motivates us to expand our analysis to  the case where chiral coatings are additionally applied to both the first and the third grating. We demonstrate that this can improve  the overall chiral selectivity of the interferometer.

 \section{All gratings coated by chiral molecules}
\begin{figure}
\includegraphics[clip,width=1\columnwidth]{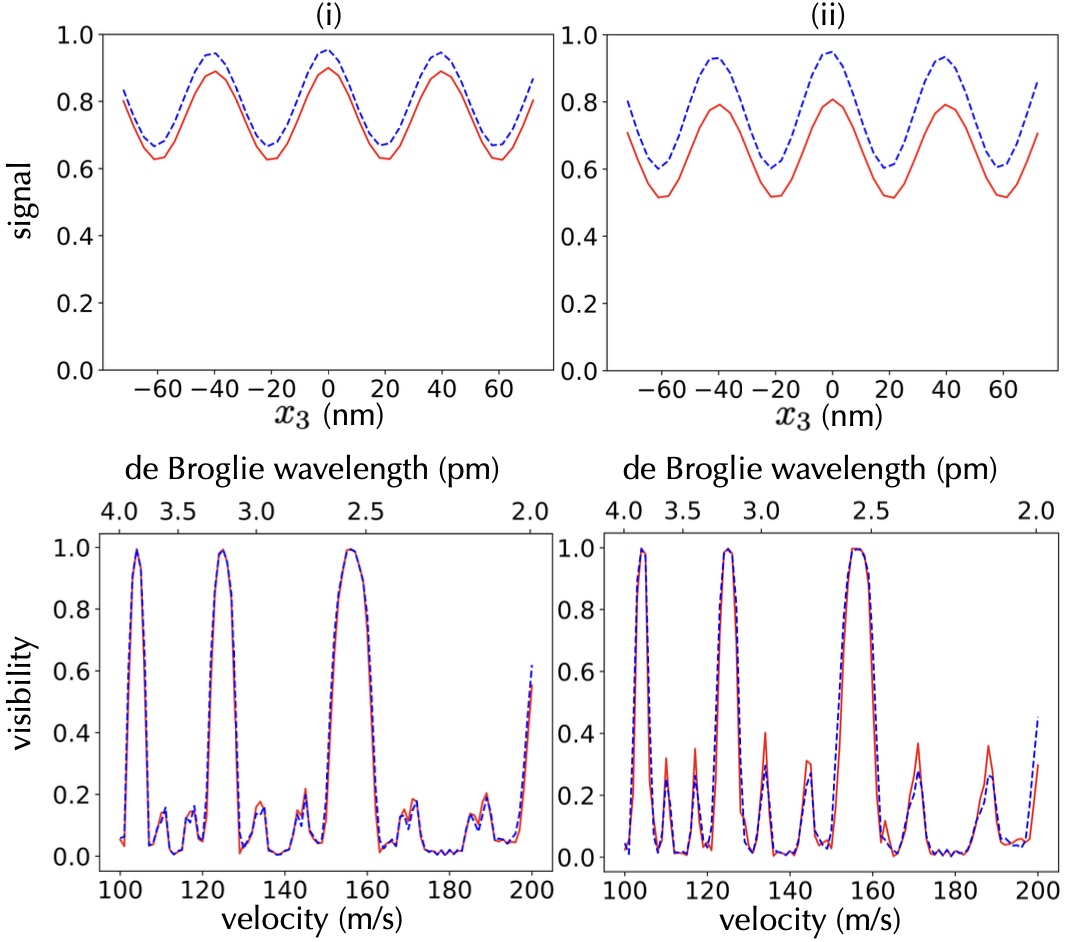} 
\caption{Same displays as in Fig.\,3, but here all gratings are identical  non-chiral SiN$_x$ masks, coated with a layer of chiral molecules. }
\label{fig4}
\end{figure}
Here, we assume to have three identical SiN gratings, all coated with a  layer of enantiomer pure, right-handed ($R_{01}^B>0$)  molecules, while all other parameters are the same as in the previous section. As before, the molecules see a combination of electric, magnetic, and chiral CP forces from the coating, as well as electric CP forces from the grating, resulting in the interferogram and fringe visibility as illustrated in Fig. \ref{fig4}. The panel (i) is simulated for  molecules with $|R_{01}|=1000\times 10^{-40}\,$cgs, $g_e=0.2$ and $g_m=5$, while panel (ii) shows the result for molecules with $|R_{01}|=5000\times 10^{-40}\,$cgs, $g_e=0.3$ and $g_m=3.3$. The right-handed molecules experience a  strong attractive force because the electric, magnetic, and chiral components all share the same sign.  This leads to a reduction in the effective grating slit width and to a loss of molecules near the coatings. Fig. 4  thus demonstrates that a coating of chiral molecules to all gratings can yield an observable  effect of chiral CP forces in the transmission signal.
Values close to our simulation parameters can be obtained for tailored molecules, such as [4]CC \cite{isobe,isobe2}. 

To explore a comprehensive range of potential test molecules, we plot in the upper panel of Fig. 5  
the difference in mean interferometer \textit{transmission signals } 
\begin{equation}
    \Delta S=\frac{1}{4fd}\int_{-2fd}^{2fd} (S^{L}(x_3)-S^{R}(x_3))/S^{L}(x_3) \,dx_3.
\end{equation}
as a function  of the  rotatory strength $R_{01}$ and the electric  anisotropy factor $g_e$.
Here,  $S_L$ and $S_R$ are the transmission signals of the left-handed and right-handed molecules, respectively.
The lower panel of Fig. 5  shows the corresponding maximum difference in fringe \textit{visibility} 
$\Delta\mathcal{V}_{\rm max}.$ Our simulations capture the maximum values across a molecular velocity range from $v_z=100\,$m/s to $200\,$m/s, and they allow for an experimental velocity resolution of  $\Delta v_z=10\,$m/s. 
This provides clear guidance for possible experiments, as soon as molecular parameters of a given species are exactly known.

Note that the magnetic component of the CP force is generally negligible compared to the electric and the chiral components of the CP force. Here we assume that $m=1000\,$Da and $\omega_{1}=2\pi \times 10^{15}\,$s$^{-1}$. The effects of the chiral CP force  are more best seen in the transmission signal but also apparent in the fringe visibility.  A particularly distinct difference between enantiomers becomes apparent for molecules with a rotatory strength $>10^{-37}\,$cgs and $g_e>0.1$. In the existing literature, it has been reported that molecules such as [4]CC can demonstrate a rotatory strength exceeding $10^{-37}\,$cgs \cite{isobe,isobe2, ref5}, and molecules such as 3-methyl-cyclopentanone  exhibit $g_e$ values exceeding 0.1 \cite{3mcp}. We anticipate that data pertaining to these parameters for various molecules will become more accessible in the future.

\begin{figure}
\includegraphics[clip,width=1\columnwidth]{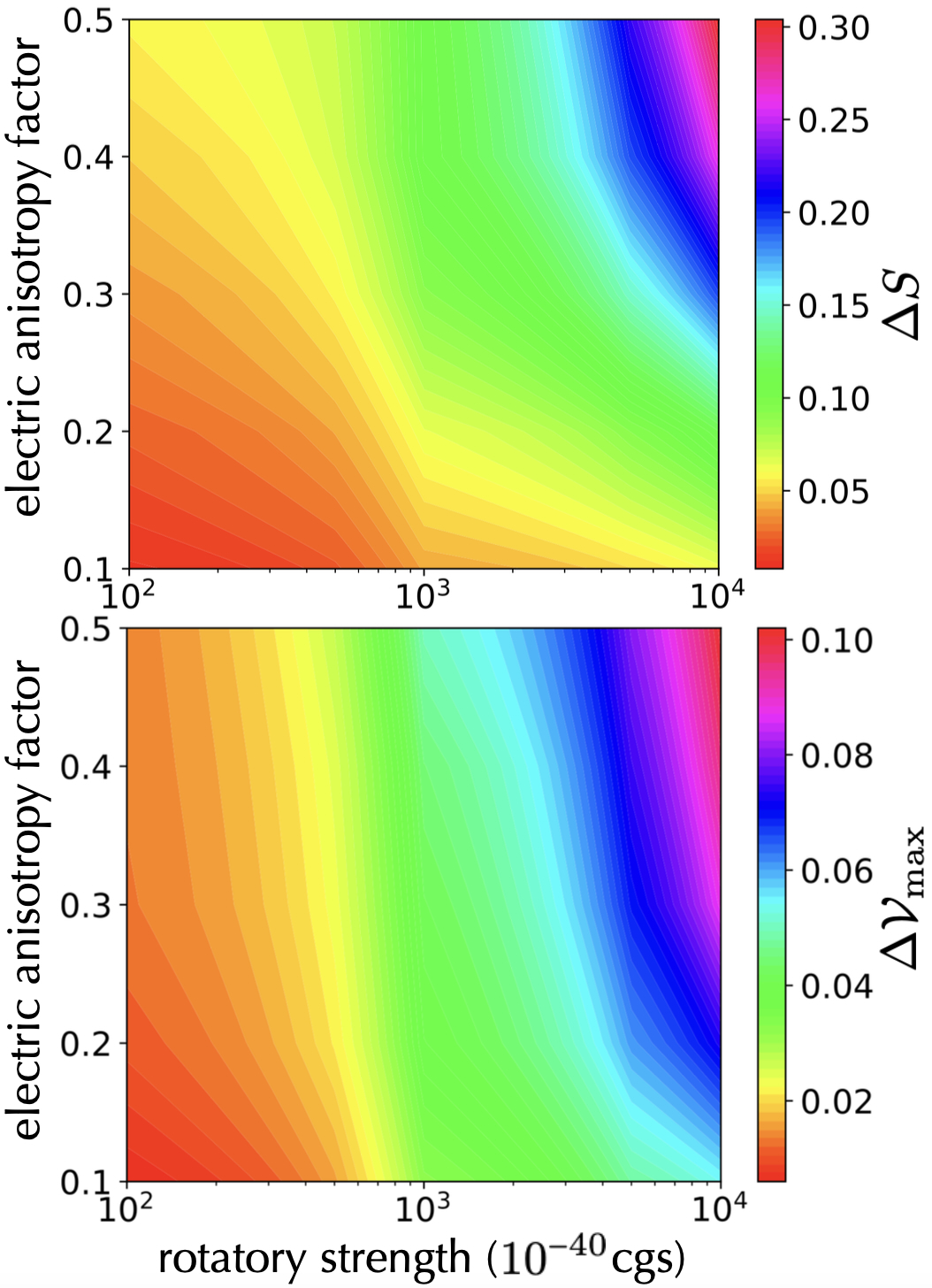} 
\caption{The maximum difference in the mean transmission signal (upper panel) and the maximum difference in fringe visibility (lower panel) between the two enantiomers are plotted as a function of the molecular rotatory strength and its electric anisotropy factor $g_e$. The maximum is found across all velocities from 100\,m/s to 200\,m/s, averaging over a resolution of 10\,m/s.}
\label{fig5}
\end{figure}

\section{Experimental aspects}
Talbot-Lau interferometers with nanomechanical gratings have been realized with electrons~\cite{McMorran2009}, positrons~\cite{Ariga2019}, atoms~\cite{Clauser1994} and molecules~\cite{Brezger2002, Juffmann2009}, before. The interferometer technology is demanding but well under control. We therefore focus on the specific requirements for the observation of chiral dispersion forces. This requires the realization of chiral gratings and beams of molecules with large optical rotatory strength such as [4]CC.

Gratings could be coated by self-assembled monolayers of chiral substances. This may be suitable for narrow gratings and one can even orient all molecules within such a layer~\cite{SAM, SAM2, SAM3}. For larger and mechanically robust structures, dip coating is a feasible technology. For structures in the 100\,nm range, however, surface tension is a challenge in all wet-chemical processes with nanostructures and needs to be addressed~\cite{Robertson-Tessi2006}. We also need many mono-layers for the proposed work.
Vapor deposition \cite{vapor,vapor2} is conceivable for hexahelicene and  3-methylyclopentanone but still needs to be experimentally verified for [4]CC. We also assume the absence of local surface charges, which may dephase matter-waves of polar molecules in transit through a nanograting~\cite{Knobloch2017}. While several highly chiral molecules, such as [4]CC, are not expected to have a permanent electric dipole moment in their thermal ground state, molecular modelling suggests that conformers with dipole moments up to 5\,Debye may transiently develop on the time scale of nanoseconds. This can be suppressed by launching the molecules into a cryogenic buffer gas to freeze out all vibrational dynamics~\cite{Piskorski2014}.  

\section{Conclusion}
We have examined three scenarios in which chiral CP forces can influence the transmission signal and the visibility  of chiral molecules behind a Talbot-Lau matter-wave interferometer with nanogratings. When the second grating consists of a perfectly chiral nanostructure ($r_c \rightarrow 1$, $r\rightarrow0$), a small  chiral effect should be observed even for the realistic case of hexahelicene and a very pronounced effect would be expected if we found a molecule with ten times larger rotatory strength (Fig. 2). Since such nanochiral materials are experimentally demanding, we have considered the case where the second grating $G_2$ is coated with highly chiral substances. We have identified examples of molecular rotatory strengths and asymmetry factors that would make an observable difference in molecular transmission but only a small difference in matter-wave fringe visibility (Fig. 3).  
The effect can be enhanced by applying the chiral coating to all gratings (Fig. 4) since even gratings that do not contribute to the interferometer phase shift influence the transmission and contrast by an effective slit narrowing. In this case, effects become visible already when both the coating molecules and matter-wave molecules possess a rotatory strength $>10^{-37}\,$cgs and an electric anisotropy factor $g_e > 0.1$ (Fig. 4). Again the transmission is more affected than the fringe visibility. Finally, we have analyzed interference figures for the same experimental geometry but for molecules now varying vastly in their rotatory strength ($10^{-38}-10^{-36}$\,cgs), in their electric anisotropy factor ($0.1-0.5$), and samples from an experimentally plausible velocity range ($100 - 200$\,m/s) with finite band width (10\,m/s). Figure 5 thus reveals the influence of these molecular parameters and while differences in transmission and fringe visibility of 1\% may still be resolved, effects on the level of 10\% would be clearly accessible. The molecular parameters are certainly challenging, but not unfeasible and the proposed experiment may allow probing for the first time the effect of \textit{static chiral} Casimir Polder forces on isolated and here even delocalized molecules.

\begin{acknowledgements}
We thank T. Momose, H. Isobe, K. Hornberger, B. A. Stickler, S. Y. Buhmann, O. Bokareva, K. Simonovich, R. Ferstl, Y. K. Kato, A. Forde  and W. H. Zurek for helpful discussions. F.S. acknowledges support from the Los Alamos National Laboratory LDRD program under project number 20230049DR and the Center for Nonlinear Studies.
\end{acknowledgements}


\begin{thebibliography}{10}
\bibitem{Casimir1948} H. B. G Casimir and D. Polder, The influence of retardation on the London-van der Waals Forces, Phys. Rev. \textbf{73}, 360 (1948).
\bibitem{chiral}R. Zhao, J. Zhou, Th. Koschny, E. N. Economou, C. M. Soukoulis, Repulsive Casimir force in chiral metamaterials, Phys. Rev. Lett. \textbf{103}, 103602 (2009).
\bibitem{chiral2} Q. Jiang and F. Wilczek, Chiral Casimir forces: Repulsive, enhanced, tunable, Phys. Rev. B \textbf{99}, 125403 (2019).
\bibitem{chiral3}D. T. Butcher, S. Y. Buhmann, and S. Scheel, Casimir–Polder forces between chiral objects, New J. Phys. \textbf{14}, 113013 (2012).
\bibitem{chiral4}P. Barcellona, H. Safari, A. Salam, and S. Y. Buhmann, Enhanced chiral discriminatory van der Waals Interactions mediated by chiral surfaces, Phys. Rev. Lett. \textbf{118}, 193401  (2017).
\bibitem{suzuki}F. Suzuki, T. Momose, and S. Y. Buhmann, Stern-Gerlach separator of chiral enantiomers based on the Casimir-Polder potential, Phys. Rev. A \textbf{99}, 012513 (2019).
\bibitem{chiral5}H. Safari, P. Barcellona, S. Y. Buhmann, and A. Salam, Medium-assisted van der Waals dispersion interactions involving chiral molecules, New J. Phys. \textbf{22}, 053049 (2020).
\bibitem{hund}J. Trost and K. Hornberger, Hund’s Paradox and the collisional stabilization of chiral molecules, Phys. Rev. Lett. \textbf{103}, 023202 (2009).
\bibitem{hund2}B. A. Stickler, M. Diekmann, R. Berger, and D. Wang, Enantiomer superpositions from matter-wave interference of chiral molecules, Phys. Rev. X \textbf{11}, 031056 (2021).
\bibitem{hund3}W. H. Zurek, {\it Decoherence and Quantum Darwinism: From Quantum Foundations to Objective Classical Reality} (Cambridge Univ. Press, 2024), to be published.
\bibitem{homo} A. Guijarro and M. Yus, {\it The Origin of Chirality in the Molecules of Life} (Royal Society of Chemistry, Cambridge, 2009).
\bibitem{CISS}  R. Naaman, Y. Paltiel, and D. H. Waldeck, Chiral induced spin selectivity gives a new twist on spin-control in chemistry, Acc. Chem. Res. \textbf{53}, 2659 (2020).
\bibitem{CISS2}A. Ghazaryan, M. Lemeshko, and A. G. Volosniev, Filtering spins by scattering from a lattice of point magnets, Commun. Phys. \textbf{3}, 178 (2020).
\bibitem{chirallight}X. Li et al., Proximity-induced chiral quantum light generation in strain-engineered WSe2/NiPS3 heterostructures, Nature Materials \textbf{22}, 1311 (2023).
\bibitem{ref3}A. Forde, D. Ghosh, D. Kilin, A. C. Evans, S. Tretiak and A. J. Neukirch, Induced chirality in halide perovskite clusters through surface chemistry, J. Phys. Chem. Lett. \textbf{13}, 686 (2022).
\bibitem{ref} A. Yokoyama, M. Yoshida, A. Ishii, and Y. K. Kato, Giant circular dichroism in individual carbon nanotubes induced by extrinsic chirality, Phys. Rev. X \textbf{4}, 011005 (2014).
\bibitem{ac} C. Ye, Q. Zhang, Y.-Y. Chen, and Y. Li, Determination of enantiomeric excess with chirality-dependent ac Stark effects in cyclic three-level models, Phys. Rev. A \textbf{100}, 033411 (2019).
\bibitem{energy}J. C. Franz, S. Y. Buhmann, and A. Salam, Macroscopic quantum electrodynamics theory of resonance energy transfer involving chiral molecules, Phys. Rev. A \textbf{107}, 032809 (2023).
\bibitem{optical}R. P Cameron, S. M. Barnett, and A. M. Yao, Discriminatory optical force for chiral molecules, New J. Phys. \textbf{16}, 013020 (2014).
\bibitem{optical2}N. Kravets, A. Aleksanyan, and E. Brasselet, Chiral optical Stern-Gerlach Newtonian experiment, Phys. Rev. Lett. \textbf{122}, 024301  (2019).
\bibitem{optical3}I. Tutunnikov, E. Gershnabel, S. Gold, and I. S. Averbukh, Selective orientation of chiral molecules by laser fields with twisted polarization, J. Phys. Chem. Lett. \textbf{9}, 1105  (2018).
\bibitem{optical4}A. A. Milner et al., Controlled enantioselective orientation of chiral molecules with an optical centrifuge, Phys. Rev. Lett. \textbf{122}, 223201  (2019).
\bibitem{optical5} C. Genet, Chiral light–chiral matter interactions: an optical force perspective, ACS Photonics \textbf{9}, 319 (2022).
\bibitem{Grisenti1999} R. E. Grisenti, W. Schöllkopf, J. P. Toennies, G. C. Hegerfeldt, and T. Köhler, Determination of atom-surface van der Waals potentials from transmission-grating diffraction intensities, Phys. Rev. Lett. \textbf{83}, 1755-1758 (1999).
\bibitem{Bruehl2002} R. Brühl, P. Fouquet, R. E. Grisenti, J. P. Toennies, G. C. Hegerfeldt, T. Köhler, M.Stoll, and C. Walter, The van der Waals potential between metastable atoms and solid surfaces: Novel diffraction experiments vs. theory, Europhys. Lett.  \textbf{59}, 357-363 (2002).
\bibitem{Lonij2010} V. P. A. Lonij, C. E. Klauss, W. F. Holmgren,  and A. D. Cronin, Atom diffraction reveals the impact of atomic core electrons on atom-surface potentials, Phys. Rev. Lett. \textbf{105}, 233202, (2010).
\bibitem{Garcion2021}	C. Garcion, N. Fabre, H. Bricha, F. Perales, S. Scheel, M. Ducloy, and G. Dutier, Intermediate-range Casimir-Polder interaction probed by high-order slow atom diffraction, Phys. Rev. Lett. \textbf{127}, 170402 (2021).
\bibitem{Bouton2023} Q. Bouton, G. Dutier, N. Fabre, E. Charron, C. Garcion, N. Gaaloul, and L. Julien, Quantum description of atomic diffraction by material nanostructures, arXiv:2312.12818v1 (2023).
\bibitem{Arndt1999} M. Arndt, O. Nairz, J. Voss-Andreae, C. Keller, G. van der Zouw and A. Zeilinger, Wave–particle duality of C60 molecules, Nature \textbf{401}, 680-682 (1999).
\bibitem{Brezger2002} B. Brezger, L.  Hackermüller, S. Uttenthaler, J.  Petschinka, M.  Arndt, and A. Zeilinger, Matter-wave interferometer for large molecules, Phys.  Rev.  Lett. \textbf{88}, 100404 (2002).
\bibitem{Brand2015}C. Brand, M. Sclafani, C. Knobloch, Y. Lilach, T. Juffmann, J. Kotakoski, C. Mangler, A. Winter, A. Turchanin, J. Meyer, O. Cheshnovsky and M. Arndt, An atomically thin matter-wave beamsplitter, Nat. Nanotechn.
\textbf{10}, 845 (2015)
\bibitem{Knobloch2017} C. Knobloch, B. A. Stickler, C. Brand, M. Sclafani, Y. Lilach, T. Juffmann, O. Cheshnovsky, K. Hornberger, M. Arndt, On the role of the electric dipole moment in the diffraction of biomolecules at nanomechanical gratings, Fortschr. Phys. \textbf{65}, 1600025 (2017).
\bibitem{Juffmann2009}	T. Juffmann, S. Truppe, P. Geyer, A. G. Major, S. Deachapunya, H. Ulbricht, and M. Arndt, Wave and particle in molecular interference lithography, Phys. Rev. Lett. \textbf{103}, 263601 (2009).
\bibitem{klaus}K. Hornberger, J. E. Sipe, and M. Arndt, Theory of decoherence in a matter wave Talbot-Lau interferometer, Phys. Rev. A \textbf{70}, 053608 (2004).


\bibitem{Fein2020}  Y.Y. Fein,  F. Kiałka, P. Geyer, S. Gerlich, M.  Arndt,  Coriolis compensation via gravity in a matter-wave interferometer, New J. Phys. \textbf{22}, 033013,  (2020)
\bibitem{Nimmrichter2011}	S.  Nimmrichter, P. Haslinger, K. Hornberger, and M. Arndt, Concept of an ionizing time-domain matter-wave interferometer, New J. Phys. \textbf{13}, 075002 (2011).

\bibitem{Juffmann2012} T. Juffmann, A. Milic, M. Müllneritsch, P. Asenbaum, A. Tsukernik, J. T\"{u}xen, M. Mayor, O. Cheshnovsky, and M. Arndt, Real-time single-molecule imaging of quantum interference, Nature Nanotechn. \textbf{7}, 297 (2012).
\bibitem{Hackermuller2003}
L. Hackermüller, S. Uttenthaler, K. Hornberger, E. Reiger, B. Brezger, A. Zeilinger, and M. Arndt, Wave nature of biomolecules and fluorofullerenes, Phys. Rev. Lett. \textbf{91}, 090408 (2003).
\bibitem{Strauss2023}
M. Strauß, A. Shayeghi, M.  F. X. Mauser
P. Geyer, T. Kostersitz, J. Salapa, O. Dobrovolskiy
S. Daly, J. Commandeur, Y. Hua, V. Köhler
M. Mayor, J. Benserhir, C. Bruschini, E. Charbon
M. Castaneda, M. Gevers, R. Gourgues, N. Kalhor, A. Fognini, M. Arndt, Highly sensitive single-molecule detection of macromolecule ion beams, Sci. Adv. \textbf{9}, (2023)
\bibitem{isotropic}J. Lekner, Optical properties of isotropic chiral media, Pure Appl. Opt.  \textbf{5}, 417 (1996).
\bibitem{scheel}S. Scheel and S. Buhmann, Macroscopic quantum electrodynamics - concepts and applications, Acta Phys. Slovaca \textbf{58}, 675 (2008).
\bibitem{stefan} S. Y. Buhmann, {\it Dispersion Forces I: Macroscopic
Quantum Electrodynamics and Ground-State Casimir, Casimir-Polder and van der Waals Forces} (Springer, Berlin, 2012).
\bibitem{green}S. M. Ali, T. M. Habashy and J. A. Kong, Spectral-domain dyadic Green’s function in layered chiral media, J. Opt. Soc. Am. A \textbf{9}, 413 (1992).
\bibitem{nano} C. Brand, J. Fiedler, T. Juffmann, M. Sclafani, C. Knobloch,
S. Scheel, Y. Lilach, O. Cheshnovsky, and M. Arndt, A Green's function approach to modeling molecular diffraction in the limit of ultra-thin gratings, Annalen der Physik  \textbf{527}, 580 (2015).
\bibitem{Simonovic2024}
K. Simonovic, R. Ferstl, A.  Barlow, A. Shayeghi, c. Brand, M. Arndt,
Impact of molecular properties on diffraction at nanomasks with low charge density, arXiv:2401.05854v1 (2024).
\bibitem{cutoff}S. Nimmrichter and K. Hornberger, Theory of near-field matter-wave interference beyond the eikonal approximation, Phys. Rev. A \textbf{78}, 023612 (2008).
\bibitem{cutoff2} J. Fiedler and B. Holst, An atom passing through a hole in a dielectric membrane: impact of dispersion forces on mask-based matter-wave lithography, J. Phys. B \textbf{55 } 025401 (2022). 
\bibitem{hexadata}	I. Warnke and F. Furche, Circular dichroism: electronic, WIREs Comput. Mol. Sci.\textbf{ 2}, 150 (2012).
\bibitem{ani} J. A. Schellman, Circular dichroism and optical rotation, Chem. Rev. \textbf{75}, 323 (1975).
\bibitem{diego}A. P. McCauley et al., Microstructure effects for Casimir forces in chiral metamaterials, Phys. Rev. B \textbf{82}, 165108  (2010).
\bibitem{andrews} D. L. Andrews, D. P. Craig, and T. Thirunamachandran, Molecular quantum electrodynamics in chemical physics, Int. Rev. Phys. Chem. \textbf{8}, 339-383 (1989).
\bibitem{salam} A. Salam, {\it Molecular Quantum Electrodynamics: Long-Range Intermolecular Interactions} (Wiley, Hoboken, NJ, 2010).
\bibitem{SiN}S. Y. Buhmann, S. Scheel, S. Å. Ellingsen, K. Hornberger, and A. Jacob, Casimir-Polder interaction of fullerene molecules with surfaces, Phys. Rev. A \textbf{85}, 042513 (2012).
\bibitem{isobe}S. Sato, A. Yoshii, S. Takahashi, S. Furumi, M. Takeuchi, and H. Isobe, Chiral intertwined spirals and magnetic transition dipole moments dictated by cylinder helicity, Proc. Natl. Acad. Sci. U. S. A. \textbf{114}, 13097 (2017).
\bibitem{isobe2}K. Kogashi, T. Matsuno, S. Sato, and  H. Isobe, Narrowing segments of helical carbon nanotubes with curved aromatic panels, Angew. Chem. Int. Ed. \textbf{58}, 7385 (2019).
\bibitem{ref5} Y. Onaka, S. Tanaka, A. Kobayashi, T. Matsuno, and H. Isobe, A large-bore chiral cylindrical molecule prone to radial deformations, Tetrahedron Lett. \textbf{96}, 153774 (2022).
\bibitem{3mcp}  D. Kr\"{o}ner, Chiral distinction by ultrashort laser pulses: electron wavepacket dynamics incorporating magnetic interactions, J. Phys. Chem. A \textbf{115}, 14510-14518 (2011).
\bibitem{McMorran2009}	B. J. McMorran and A. D. Cronin, An electron Talbot interferometer, New J.  Phys. \textbf{11}, 033021 (2009).
\bibitem{Ariga2019} S. Sala, A.  Ariga, A.  Ereditato, R.  Ferragut, M.  Giammarchi, M.  Leone, C.  Pistillo,  and P.  Scampoli, First demonstration of antimatter wave interferometry, Science Adv. \textbf{5}, eaav7610 (2019).
\bibitem{Clauser1994} J. F. Clauser and S. Li, Talbot-vonLau atom interferometry with cold slow potassium, Phys. Rev. A \textbf{49}, R2213 (1994).
\bibitem{SAM} J. C. Love, L. A. Estroff, J. K. Kriebel, R. G. Nuzzo, and G. M. Whitesides, Self-assembled monolayers of thiolates on metals as a form of nanotechnology, Chem. Rev. \textbf{105}, 1103 (2005).
\bibitem{SAM2} P. V. M\"{o}llers, S. Ulku, D. Jayarathna, F. Tassinari, D. Nürenberg, R. Naaman, C. Achim, and H. Zacharias, Spin-selective electron transmission through self-assembled monolayers of double-stranded peptide nucleic acid, Chirality \textbf{33}, 93 (2021).
\bibitem{SAM3} B. G\"{o}hler, V. Hamelbeck, T. Z. Markus, M. Kettner, G. F. Hanne, Z. Vager, R. Naaman, and H. Zacharias, Spin selectivity in electron transmission through self-assembled monolayers of double-stranded DNA, Science \textbf{331}, 894 (2011).
\bibitem{Robertson-Tessi2006} M. Robertson-Tessi, R. J. Wild, A. D. Cronin, and Tim Savas, Cleaning silicon nitride gratings with liquid immersion, J. Vac. Sci.Tech. B \textbf{24}, 1409  (2006).
\bibitem{vapor} P. M. Martin, {\it Handbook of Deposition Technologies for Films and Coatings}  (Elsevier, Norwich, NY, 2009).
\bibitem{vapor2}H. Al-Bustami, S. Khaldi, O. Shoseyov, S. Yochelis, K. Killi, I. Berg, E. Gross, Y. Paltiel, and R. Yerushalmi, Atomic and Molecular Layer Deposition of Chiral Thin films showing up to 99\% spin selective transport, Nano Lett. \textbf{22}, 5022  (2022).
\bibitem{Piskorski2014}
J. Piskorski, D. Patterson, S. Eibenberger, and J. M. Doyle, Cooling, spectroscopy and non-Sticking of trans-Stilbene and Nile red, ChemPhysChem \textbf{15}, 3800 (2014).


\end{thebibliography}
\end{document}